\documentclass[aps,pre,onecolumn]{revtex4-2}
\usepackage{amsmath,amsfonts,amssymb,bm,graphicx}
\usepackage{textcomp}%
\usepackage{subcaption}

\begin{document}

\title{Time arrow in open-boundary one-dimensional stochastic dynamics: effects of discrete thermal kicks over temperature interface}


\author{Chi-Lun Lee$^1$}
\email[]{lee.chilun@gmail.com}
\author{Yu-Syuan Lin$^1$}
\author{Pik-Yin Lai$^{1,2}$}
\email[]{pylai@phy.ncu.edu.tw}
\affiliation{$^1$Department of Physics and Center for Complex Systems, National Central University, Chung-Li District, Taoyuan City 320, Taiwan, R.O.C.}
\affiliation{$^2$Physics Division, National Center for Theoretical Sciences, Taipei 10617, Taiwan, R.O.C.}

\date{\today}

\begin{abstract}
Does nonequilibrium characteristics exhibit in one-dimensional open-boundary stochastic systems?  We introduce a discrete-timestep stochastic model, where the particle dynamics follows a discrete overdamped stochastic equation, and thermal kicks are presented over discrete timesteps.  These thermal kicks are derived from a position-dependent temperature profile with a sharp interface.  The discrete characteristic and the It\^{o}-type stochastic scheme make our model distinct from the conventional discussions of an overdamped colloidal model.  Under the open-boundary condition, one cannot observe any net probability current in the nonequilibrium steady state (NESS).  On the other hand, the nonequilibrium nature of this system can be extracted through the asymmetry in forward and backward transition probabilities, as is reported in this work through the simulation and theoretic analysis.  In particular, our result shows prominent time irreversibility near the temperature interface.  We propose a hidden-gyration scenario that demonstrates the source of irreversibility, while the collapse of gyrations on the one-dimensional coordinate accounts for the absence of probability current. 
\end{abstract}

\maketitle

\section{Introduction}
\label{sec_intro}
In the discussions of stochastic dynamics, the systems of autonomous gyrators\cite{Filliger_PRL2007, Imparato_gyrator, harmonic_analytic2013, harmonic_analytic2018, exp_gyrator2017, RC_gyrator_2017,chang2021} have recently drawn great interest.  For these systems, the most renowned feature is the emergence of a perpetual, average circulating motion, despite the fact that the dynamics is highly random within short-time intervals.  In two-dimensional stochastic systems, such gyrating behavior can arise due to the asymmetry in thermal fluctuations along the two coordinates\cite{Filliger_PRL2007,exp_gyrator2017,RC_gyrator_2017}, or due to the temperature inhomogeneity in space\cite{sasa2000,Sekimoto_book}.  The understanding of such autonomous gyrating systems is crucial for the development of microscopic heat engines.

In one-dimensional systems, persistent average gyration has been studied since the famous B\"uttiker-Landauer discussions in the 1980s\cite{Buttiker, Landauer}.  Through the closed-loop topology, a nonvanishing probability current can be achieved through a proper setting of temperature and potential profiles.  The directed average dynamics, as an indicator of time irreversibility, is an important characteristic of the second law in thermodynamics, or its recent variants in stochastic dynamics, the fluctuation theorems\cite{Cohen93, Lebowitz99, Crooks99, Jarzynski00, Seifert05}.  In the B\"uttiker-Landauer engine, as well as in two-dimensional Brownian gyrators, it can be shown that the average net heat dissipation along the gyrating loop is positive.  This is in agreement with the second law in the sense that the overall change in total entropy along a closed autonomous cycle is, on average, positive.

In contrast to the examples above, a steady-state probability current in position space is absent regarding the stochastic dynamics in a one-dimensional, open-boundary system.  In particular, the lack of current under a nonuniform temperature profile indicates that the nonequilibrium characteristics are less obvious.  For example, Hondou and Sekimoto\cite{Hondou00} studied the stochastic energetics of the B\"uttiker-Landauer engine.  Starting from the underdamped description following the Kramers equation, they succeeded in deriving the average heat transfer by the Brownian particle from the hot to the cold heat reservoir.  And they verified that this heat transfer did not vanish in the overdamped limit.   
In addition to the aforementioned study, the nonequilibrium nature in temperature-gradient systems has also been discussed in Refs.~\cite{Sekimoto_book,Celani12,Shiraishi_book}.  The above works again introduced the discrepancy in overdamped stochastic dynamics, as well as its shortage of contribution in total entropy dissipation of temperature-gradient systems.  Specifically, Celani {\it et al.}\cite{Celani12} obtained in underdamped stochastic dynamics an additional source of time irreversibility, or ``entropy anomaly'' in overall dissipation.  This irreversible behavior is attributed to the circulating steady-state behavior in the phase space, and such an entropy anomaly persists even if one proceeds towards the overdamped limit.

Inspired by the aforementioned successes showing the overall entropy production, and out of curiosity, we look to the possible existence of nonequilibrium characteristic in an analogous overdamped model in an open-boundary condition.  We focus on the B\"uttiker-Landauer-like scenario, of which temperature changes discontinuously.  Our system of interest is governed by a discrete overdamped stochastic equation.  This system can be mimicked by considering a particle in an overdamped medium at very low temperature in one dimension, while the particle is subject to extra external kicks that have a discontinuous-over-position thermal profile.  These ``thermal'' kicks occur over discrete time intervals, and each kick is contributed by an individual temperature bath.  The model introduced here can be suited for stochastics of a Brownian particle with additional kicks from active particles or artificial feedback control\cite{Jun1, Jun2}, or even a granular gas. Through theoretical arguments and simulation analysis on transition probabilities, we find that the nonequilibrium characteristic can be exhibited through time irreversibility.  Moreover, a ``virtual'' gyrating behavior can be found at the local scale, for transitions near the temperature discontinuity.  

The time irreversibility observed here is distinct from the discussions in entropy anomaly\cite{Celani12} and heat transfer by the Brownian particle\cite{Hondou00}, both derived from the limit of underdamped kinetics.  
Our model of study, though simple enough, can give to the irreversible feature.  And it shares a common characteristic with the previous underdamped descriptions in that the particle can penetrate to the other temperature region and thus exchange information before losing its memory.  In the previous works, it is the underdamped nature that maintains its memory, while in our current model, it is the discreteness in kicking timestep that makes it possible to travel across the interface in one thermal kick.


Our work is structured as follows: first we introduce our system of study, its dynamic stochastic equation, as well as a corresponding Fokker-Planck equation analysis that works for infinitesimal timesteps.  Then we proceed to compute the finite-timestep transition probabilities and demonstrate through a crude approximation that time reversibility, or its synonym, detailed balance, may be violated.  We go on to examine the violation of detailed balance through theoretical arguments, and the simulation result also confirms the existence of time irreversibility.  Meanwhile, the steady-state probability distribution obtained from the simulations shows deviation from the Fokker-Planck equation analysis.  We then seek to fit the result of the probability distribution to that of a corresponding continuous-time model.  This is achieved through the introduction of an effective temperature with a smoothed-out profile.  Finally, we give an estimate on the total entropy production rate following the probabilistic approach.

\section{System}
\label{sec_modelmethod}
We consider a particle moving in one dimension.  Assuming the particle to be subject to random thermal kicks that occur discretely with time interval $\Delta t$,  the stochastic dynamics of the particle is governed by the equation
\begin{equation}
	\gamma (x'-x) = F(x)\Delta t + \sqrt{2 \gamma k_{\rm B} T(x)}\, \xi \,\Delta t\, ,
	\label{eqn_Langevin2}
\end{equation}
where $x$ and $x'$ are the original and updated positions, respectively.  $\displaystyle F = -\frac{dU}{dx}$ represents the conservative force, and $\xi\, \Delta t$ corresponds to the noise in a typical Wiener process, and $\gamma$ representing the damping coefficient.  
Thus a Brownian particle at position $x$ is subject to a smooth potential $U(x)$, while the last term in the equation represents the source of the random force.  This fluctuating force is Gaussian white and uncorrelated, namely, $\langle \xi(t)\xi(t')\rangle= \delta(t-t')=\delta_{tt'}/\Delta t$.  For simplicity, all the physical quantities we use in this work are made dimensionless.  Under this representation, the Boltzmann constant and $\gamma$ are both set to be unity.  The latter can be achieved through a rescaling of time by the factor $1/\gamma$.  The temperature is profiled by $T(x) = T_H$ for $x>x_0$ and $T(x) = T_L$ for $x\leq x_0$.  Note that the product $\sqrt{T(x)}\, \xi$ in Eq.~\eqref{eqn_Langevin2} adopts It\^{o}'s scheme.  This scheme is better suited for the discussions of granular gases, and molecular-motor movements\cite{Cercignani_book,Puglisi_book,Prost97}, as well other systems with discrete kicks\cite{Jun1, Jun2}.

When the time interval between success kicks $\Delta t$ becomes negligible, Eq.~\eqref{eqn_Langevin2} reduces to the well-known overdamped Langevin equation:
\begin{equation}
	\frac{dx}{dt} = - \frac{dU}{dx} + \sqrt{2 T(x)}\, \xi \, .
	\label{eqn_Langevin}
\end{equation}

The stochastic dynamics of Eq.~\eqref{eqn_Langevin}, under the dimensionless representation, can be equivalently described via the Fokker-Planck equation (FPE):
\begin{equation}
	\frac{\partial P}{\partial t} = \frac{\partial}{\partial x} \left[ \frac{\partial (TP)}{\partial x} + P\frac{\partial U}{\partial x}\right] \, ,
	\label{eqn_FP}
\end{equation}
where $P$ is the probability density of the particle.  Note that the equivalency between the descriptions in Eqs.~\eqref{eqn_Langevin} and \eqref{eqn_FP} are based on the scenario when random kicks are continuously dispersed over time.  In the nonequilibrium steady-state (NESS) regime, one then has $\displaystyle \frac{d\, j}{dx} = 0$, while
\begin{equation}
	j = - \left[ \frac{\partial (TP)}{\partial x} + P\frac{\partial U}{\partial x}\right]
	\label{eqn_j}
\end{equation}
represents the probability current.  
At the steady state, the probability current is constant as a result of the equation of continuity.  And since the probability vanishes as $x \to \pm \infty$,  $j= 0$ everywhere in the one-dimensional open-boundary system, or a nonzero persistent current would result in a change of probability at large $|x|$.  
Therefore, within each constant-temperature domain, the probability density is proportional to the Boltzmann distribution.  To determine the prefactors at these two regions, we consider an infinitesimal transition across the temperature discontinuity $x=x_0$.  Since Eq.~\eqref{eqn_j} implies $d(TP) + PdU = 0$ in the NESS, if $PdU$ is negligible for infinitesimal transition step, so is $d(TP)=0$.  Therefore, one has
\begin{equation}
	(TP)|_{x=x_0^{-}} = (TP)|_{x=x_0^{+}} \, .
	\label{eqn_TP}
\end{equation}
Note that the discontinuous temperature profile leads to a probability density discontinuity.  Thus, according to the Fokker-Planck equation, one has
\begin{equation}
	P = {\rm{constant}} \cdot \frac{e^{-\beta(U-U_0)}}{T} \, ,
	\label{eqn_P_FP}
\end{equation}
where $\beta \equiv 1/T$, and $U_0$ is the potential energy right at the temperature discontinuity.\footnotemark[1]
\footnotetext[1]{Note that strictly speaking, Eq.~\ref{eqn_P_FP} is not the accurate solution.  When one applies the Fokker-Planck operator on it, the result is not strictly a zero function.  The rigorous version of the FPE solution is given in Eq.~\ref{eqn_prob}.  Nevertheless, both expressions are pointwise identical except at (infinitesimally near) $x=x_0$.}

\begin{figure}[h]
	\includegraphics[width=0.7\textwidth,height=0.2\textwidth]{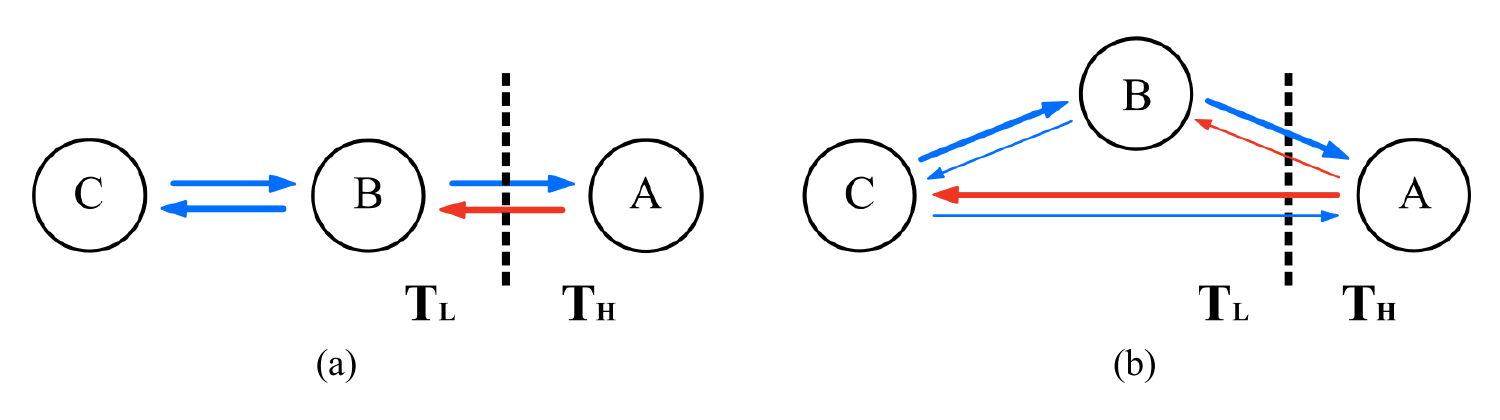}
	\caption{Illustrations about the two scenarios.  Arrows represent probability currents (width signifying magnitude).  Points $C$, $B$, $A$ represent successive discrete positions from left to right, while point $A$ is at a higher temperature.  (a) The chain-connecting scenario.  Note that the net steady-state current between nodes has to be zero everywhere (detailed balance).  (b) The loop scenario.  In finite-timestep stochastic dynamics, thanks to the large diffusion parameter at $A$, point $A$ can reach $C$ directly over one single timestep $\Delta t$.  This feature gives rise to the loop-like scenario and contributes to a ``hidden'' gyrating current at the steady state.}
	\label{Fig_illustration}
\end{figure}

Note that the FPE represents a special case of the master equation, following the truncation of the Kramers-Moyal expansion\cite{Risken, van_Kampen} up to the second order.  The legitimacy of truncation is based on the underlying assumption that transitions are local.  In particular, for 1-dim random walks, the FPE approach is valid for models with chain-like connecting structures (see Fig.~\ref{Fig_illustration}(a)).  Regarding an open-boundary system with a chain-like structure, its steady-state dynamics meets the detailed-balance criterion as long as the probability vanishes at far distances.

In the following sections, we study the one-dimensional Brownian motion under discrete-timestep thermal kicks as described in Eq.~\eqref{eqn_Langevin2}.  We observe the failure of the detailed-balance condition, while the failure is most prominent for transitions across a temperature discontinuity.  This failure is attributed to the fact that our discussed regime cannot be faithfully modeled by an open-chain structure.  The absence of an open-chain structure also indicates a possible deviation from the FPE analysis.

\section{Transition Probability; Violation of Detailed Balance}
\label{sec_tran}  
Now we look for the irreversible characteristic of our system.  Consider a transition from position $x$ to $x'$ over timestep $\Delta t$. 
From Eq.~\eqref{eqn_Langevin2}, one can derive the conditional transition probability
\begin{equation}
	P(x'|x) = \frac{1}{\sqrt{2\pi}}\cdot\frac{1}{\sqrt{2T(x)\, \Delta t}} \exp\left[ -\frac{\Delta t}{4T(x)}(\dot{x}-F(x))^2 \right],
	\label{eqn_forward}
\end{equation}
and that for the backward transition (i.e., $x' \to x$):
\begin{equation}
	P(x|x') = \frac{1}{\sqrt{2\pi}}\cdot\frac{1}{\sqrt{2T(x')\, \Delta t}} \exp\left[ -\frac{\Delta t}{4T(x')}(-\dot{x}+F(x'))^2 \right],
	\label{eqn_backward}
\end{equation}
where $\dot{x} \equiv (x'-x)/\Delta t$. 
Let us use the abbreviation that $T' \equiv T(x')$ and $F' \equiv F(x')$, while the unprimed letters $T$ and $F$ are reserved for the quantities at position $x$. Comparison of the forward and backward conditional transition probabilities then gives
\begin{align}
	\ln \frac{P(x'|x)}{P(x|x')} & = \frac12 \ln \frac{T'}{T} + \frac{\Delta t}{4}\cdot \left[ -\frac{(\dot{x}-F)^2}{T} + \frac{(\dot{x}-F')^2}{T'}\right] \nonumber \\
	& \approx \frac12 \ln \frac{T'}{T} + \frac{\Delta t}{4}\cdot \left[ \left( \frac{1}{T'}-\frac{1}{T} \right) \cdot (\dot{x}^2 + F^2) + 2\dot{x} F \left( \frac{1}{T} + \frac{1}{T'} \right) \right] \, .
	\label{eqn_ratio1}
\end{align}
Note that the approximation of $F'$ by $F$ results in errors of magnitude $O(\Delta x^2)$ and $O(\Delta t \cdot \Delta x)$ in the last two terms in Eq.~\eqref{eqn_ratio1}, respectively ($\Delta x \equiv x'-x$).  
In terms of entropy change of the environment, the first term in Eq.\eqref{eqn_ratio1} represents the thermophoretic contribution, while the last term corresponds to the heat entropy dissipation to the environment (For a constant-temperature transition $T'=T$, Eq.~\eqref{eqn_ratio1} reduces to $(F \cdot \Delta x)/ T$\cite{Seifert05}).  Meanwhile, regarding a transition across a temperature interface, the middle term in Eq.~\ref{eqn_ratio1} reveals an additional source of entropy dissipation.  The middle term is positive whenever $T'<T$ and vice versa, and it becomes significant once $\Delta x$ reaches the diffusive scale, i.e., $(\Delta x)^2/\Delta t \gtrsim O(1)$ under the dimensionless representation.  As shown below, this term serves as the source of irreversibility.

Next we demonstrate the time irreversibility through the Kolmogorov criterion\cite{Kolmogorov}: 
Let us consider three neighboring sites $A, B$ and $C$, while the temperature at site $A$ is $T_H$ and the other two sites at a lower temperature $T_L$.  The distances between $AB$ and $BC$ are equal, and we denote this distance to be $\Delta x$.  According to the Kolmogorov criterion,
\begin{equation}
	\frac{P(A,B) P(B,C) P(C,A)}{P(B,A) P(C,B) P(A,C)} = \frac{P(A|B)  P(B|C) P(C|A)}{ P(B|A) P(C|B) P(A|C)}\, ,
	\label{eqn_joint_cond}
\end{equation}
where $P(A,B)$ represents the joint probability that the random walker is at site $B$ at time $t$ and $A$ at a later time $t+\Delta t$.  The right-hand side of Eq.~\eqref{eqn_joint_cond} is known as the loop affinity in the discussion of stochastic thermodynamics\cite{Kolmogorov, Shiraishi_book}.

With Eq.~\eqref{eqn_ratio1}, it leads to
\begin{align}
	\ln \frac{P(A,B) P(B,C) P(C,A)}{P(B,A) P(C,B) P(A,C)} &\approx
	\frac{(\Delta x)^2}{4\, \Delta t}\left(\frac{1}{T_H}-\frac{1}{T_L}\right) + 0 +
	\frac{(2\Delta x)^2}{4\, \Delta t}\left(\frac{1}{T_L}-\frac{1}{T_H}\right) \nonumber \\
	& \ \ + \frac{\Delta x}{2}\left( \frac{F}{T_H}+\frac{F}{T_L} \right) + \frac{\Delta x}{2}\left( \frac{F}{T_L}+\frac{F}{T_L} \right) - \frac{2\Delta x}{2}\left( \frac{F}{T_H}+\frac{F}{T_L} \right) \nonumber \\
	&= \frac34 \frac{(\Delta x)^2}{\Delta t}\left(\frac{1}{T_L}-\frac{1}{T_H}\right)
	+ \frac{F\, \Delta x}{2}\left( \frac{1}{T_L}-\frac{1}{T_H} \right) \, .
	\label{eqn_db_violation}
\end{align}
In our work we choose $(\Delta x)^2= O(\Delta t)$, i.e., $\Delta x$ is comparable to the diffusive scale over time $\Delta t$.  In this case, the first term in the final line of Eq.~\eqref{eqn_db_violation} outweighs the second term, and one observes significant violation in the detailed-balance condition: at least one forward-to-backward ratio among the $(A,B)$, $(B,C)$, $(C,A)$ pairs is not equal to one.

While the violation of detailed balance introduced above is attributed to the temperature change over sites, one may guess that the violation is present only for transitions straight across the temperature interface.  We now proceed to argue that in fact, the detailed-balance violation can occur for transitions across constant-temperature sites.  Let us temporarily assume that for our system of consideration, the detailed-balance condition still holds for all constant-temperature transitions.  As a result, the probability current between constant-temperature neighboring sites must be zero.  Let us enlarge our sites of consideration such that the ``node'' $A'$ includes all high-temperature sites, node $B'$ includes site $B$ only, and node $C'$ includes all low-temperature sites other than site $B$.  According to the assumption, there is be no probability current between nodes $B'$ and $C'$.  However, this result is contradicting, since a three-way steady state cannot be achieved in this manner: if there exist some nonzero current between any of the two nodes, the same amount of current must circulate through the three nodes (see Fig.~\ref{Fig_illustration}(b)).  Therefore, we come to the conclusion that the detailed-balance condition must be violated between some constant-temperature nodes.

The three-way circulation in probability current described above is as a demonstration of the ``hidden'' gyration: the violation of detailed balance between successive sites suggests that the current can gyrate among local sites.  On the other hand, the net probability current, i.e., the projection of current upon the one-dimensional coordinate, must be absent in the NESS dynamics of this open-boundary system.

Finally, we remark that if the solution of FPE (Eq.~\eqref{eqn_P_FP}) is valid, one then has
\begin{align}
	\ln \frac{P(x',x)}{P(x,x')}
	& \approx \frac32 \ln \frac{T'}{T} + \frac{\Delta t}{4}\cdot \left( \frac{1}{T'}-\frac{1}{T} \right) \cdot (\dot{x}^2 + F^2) - \frac{U'-U}{2} \left( \frac{1}{T} + \frac{1}{T'} \right) -\frac{1}{T}(U-U_0) +\frac{1}{T'}(U'-U_0) \nonumber \\
	& = \frac32 \ln \frac{T'}{T} + \frac{\Delta t}{4}\cdot \left( \frac{1}{T'}-\frac{1}{T} \right) \cdot (\dot{x}^2 + F^2) + \frac12 \left( \frac{1}{T'} - \frac{1}{T} \right) \cdot (U+U'-2U_0) \, ,
	\label{eqn_ratio2}
\end{align}
which will vanish for transitions over constant-temperature sites.
But this contradicts with our prediction in the previous paragraph.  Thus one concludes that for finite-timestep stochastic dynamics, its steady-state probability distribution may deviate from Eq.~\eqref{eqn_P_FP}, as will be shown in our simulation results.

\section{Simulation results}
\label{sec_simulation}
In this work, we consider the harmonic potential $U= x^2/2$.  The temperature discontinuity is set to the position $x_0=-2$, and our numerical result was obtained via Eq.~\eqref{eqn_Langevin} using the discrete timestep $\Delta t = 10^{-2}$ under the condition $T_H = 4$ and $T_L = 2$ unless mentioned elsewhere.  The simulation was performed over a total of $10^8$ to $10^9$ timesteps to achieve sufficient sampling of the NESS behavior.  The recorded time series in position is classified into bins of width $\delta$, so that statistical distributions can be obtained. 




\subsection{Joint Transition Probability}
\label{subsec_joint}
To quantitatively assess the violation of detailed balance predicted in section~\ref{sec_tran}, we compare the asymmetry in forward and backward transition probabilities across multiple spatial points.
Specifically, we look at the joint probability $P(x',x)$ of finding the particle at position $x$ at time $t$ and position $x'$ at a later time $t + \Delta t$, and compare it with the occurrence of the backward process $P(x,x')$.  The joint probabilities are obtained via trajectory analysis.  Note that due to the coarse-graining procedure, the notions $x$ and $x'$ in $P(x',x)$ in our data analysis actually mean all possible points residing in the bins centered at $x$ and $x'$, respectively.

We first divide the spatial domain into bins of width $\delta$, and compute the ratio of forward to backward transitions between adjacent intervals:
\begin{equation}
	R(x) \equiv \frac{P(x + \delta/2, x - \delta/2)}{P(x - \delta/2 , x + \delta/2)}\, .
	\label{eqn_ratio_joint}
\end{equation}
To study the effect of time irreversibility using the above equation, we select a $\delta$ that is smaller than the average single-step diffusion distance (0.27 for a typical simulation), so that the statistics wouldn't be over coarse-grained.  Meanwhile, $\delta$ cannot be too small.  This is because single-step diffusions over the temperatures we choose exhibit minor differences on that scale, and hence the irreversibility becomes less significant.

\begin{figure}[h]
	\includegraphics[width=0.85\textwidth]{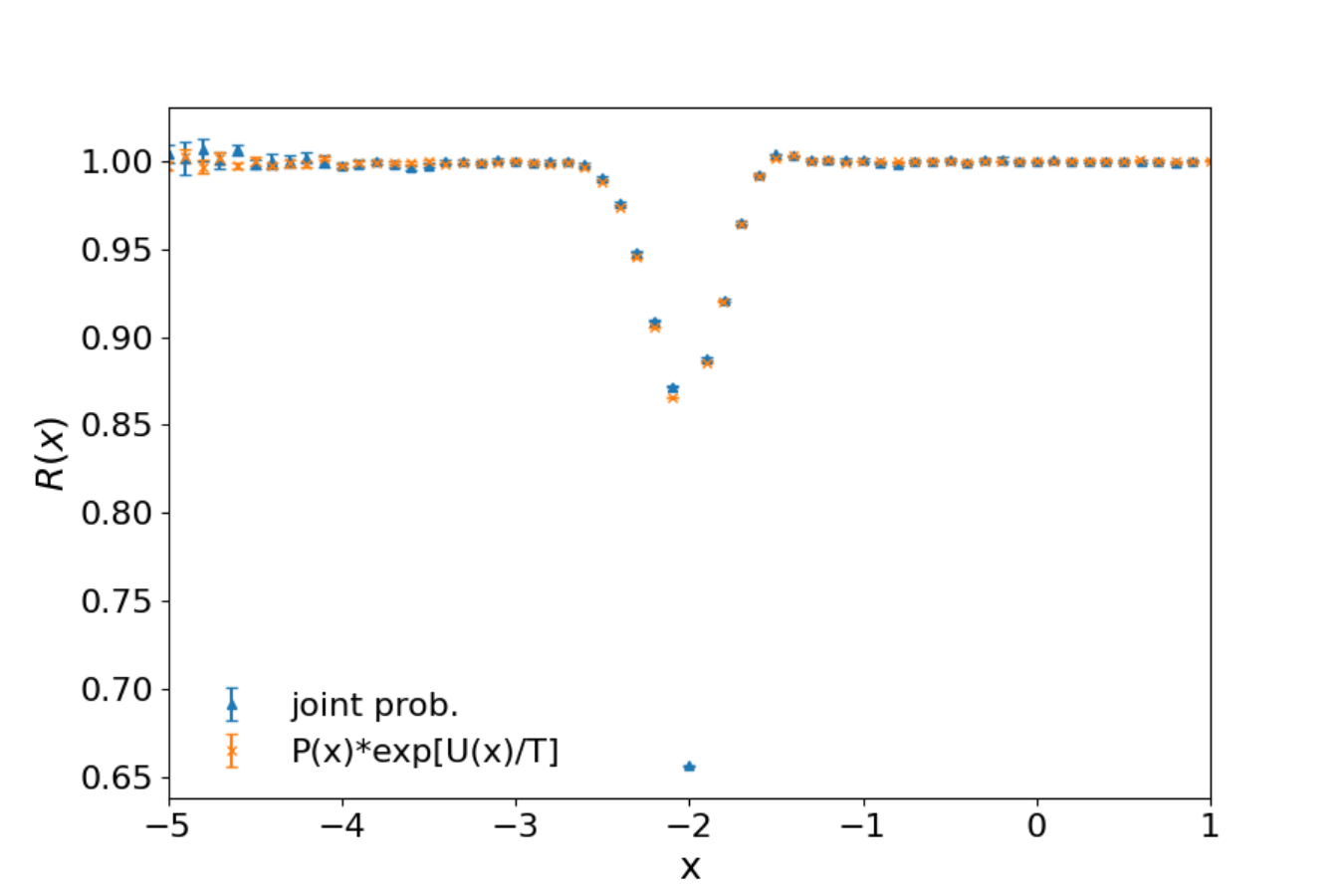}
	\includegraphics[width=0.63\textwidth]{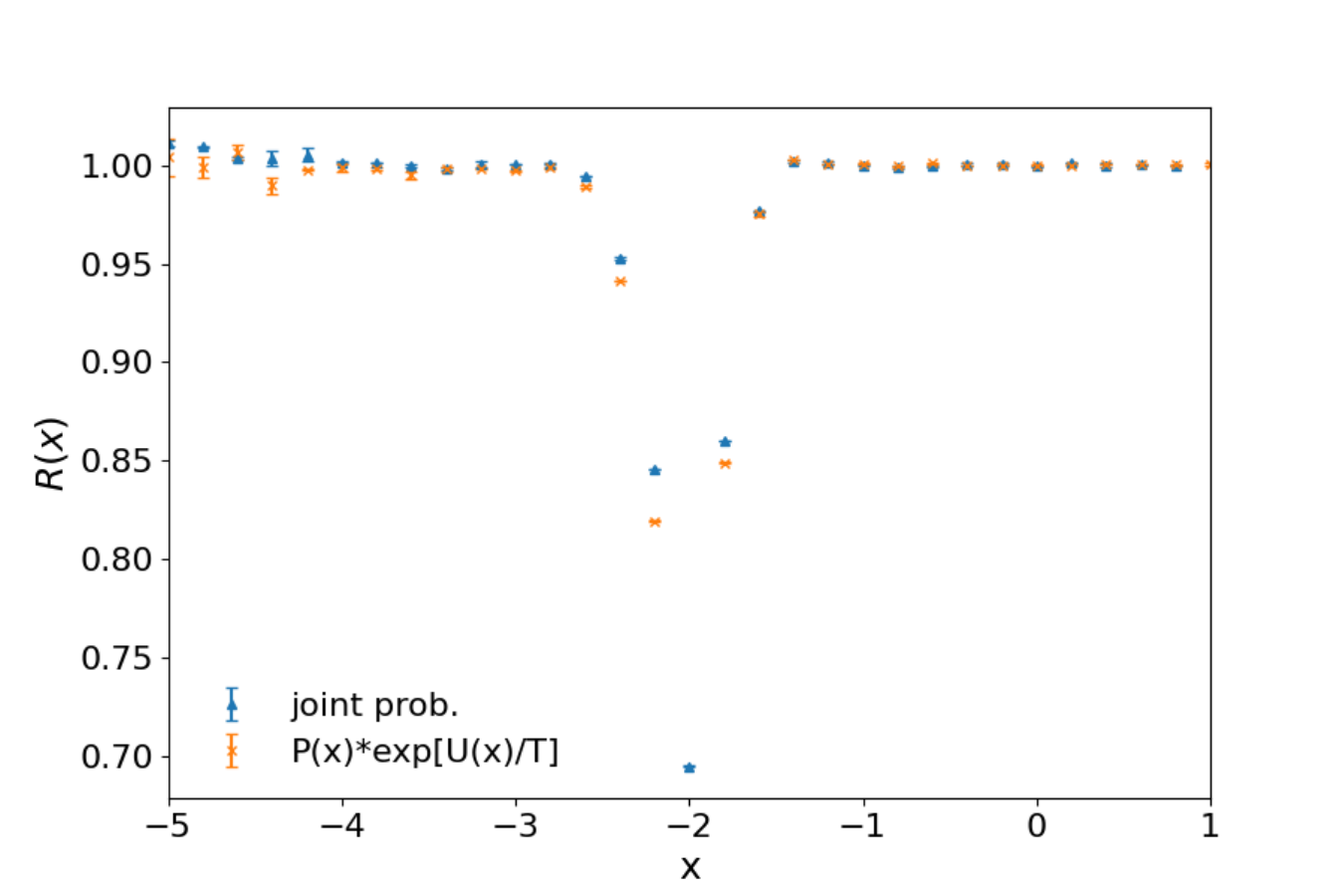}
	\includegraphics[width=0.62\textwidth]{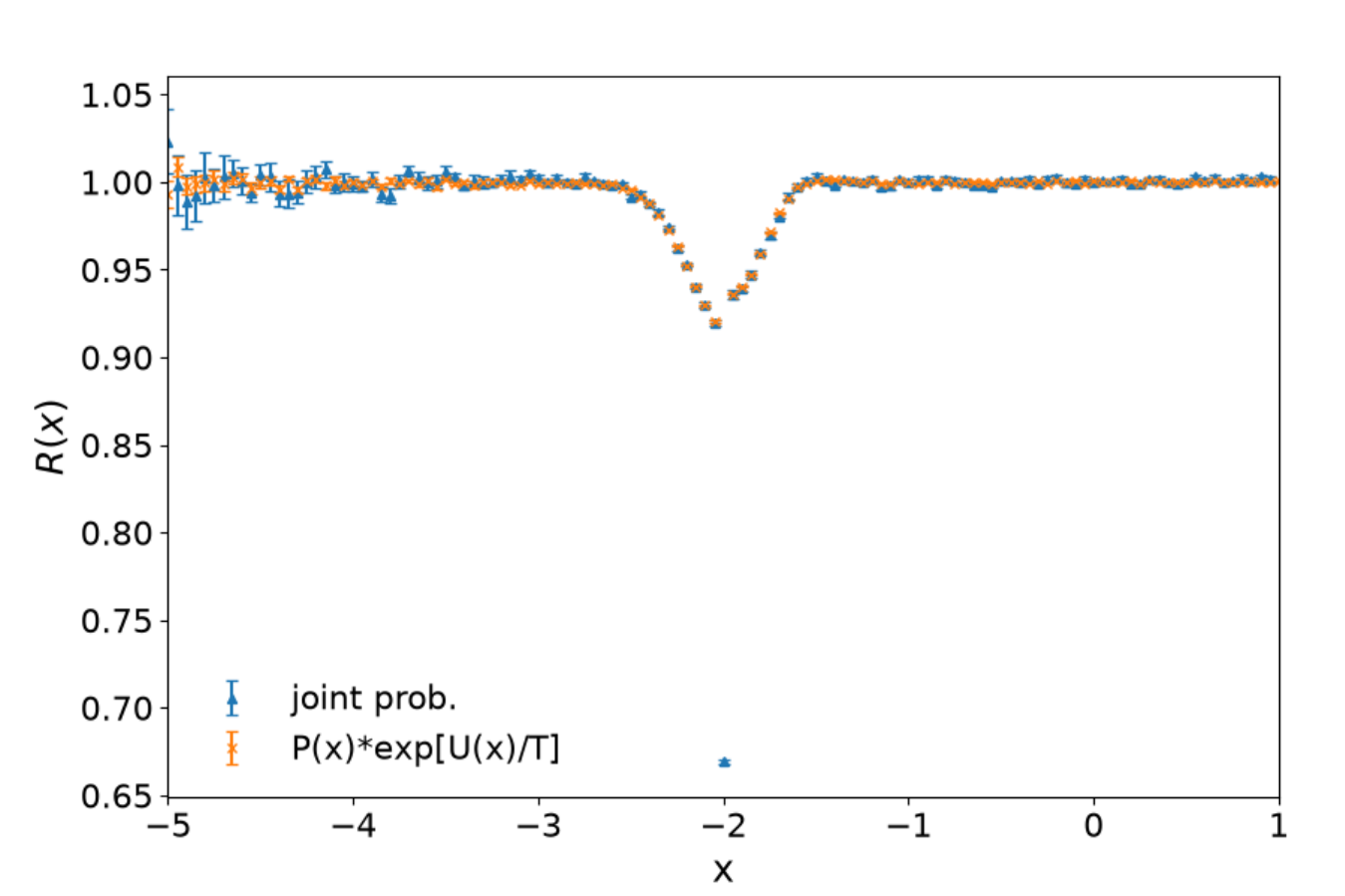}
	\includegraphics[width=0.6\textwidth]{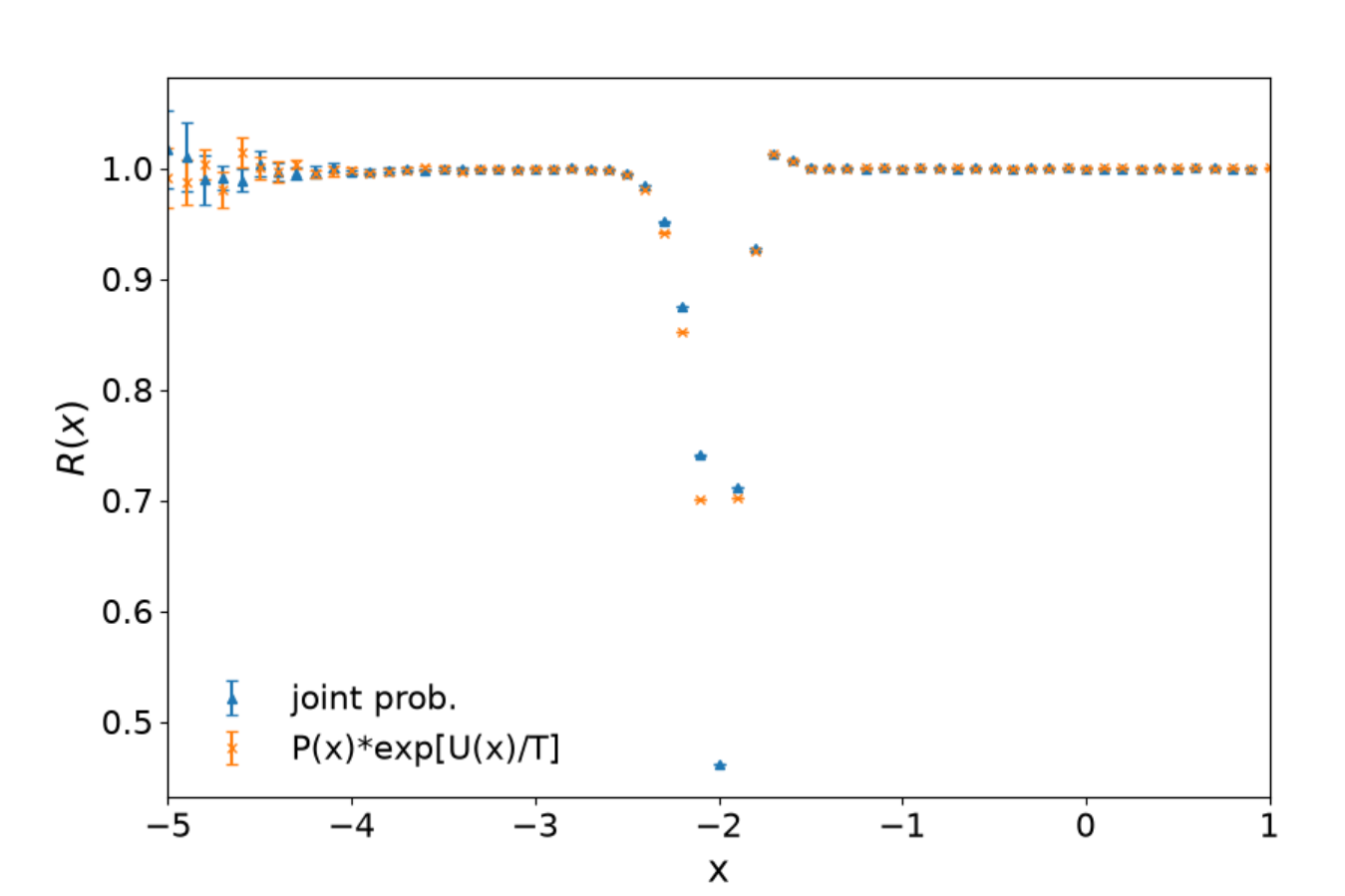}
	\caption{Ratio of joint probabilities between forward and backward transitions for (a) $T_H=4, T_L=2, \delta = 0.1$ ; (b) $T_H=4, T_L=2, \delta = 0.2$; (c) $T_H=4, T_L=2, \delta = 0.05$; (d) $T_H=4, T_L=1, \delta = 0.1$.
		The counting region width $\delta$ is determined by the global root-mean-square (RMS) displacement of the Brownian particle. Blue dots (triangle): result where joint probabilities were extracted directly from simulation data; orange dots (cross): result evaluated from Eq.~\eqref{eqn_ratio_predicted}, via the NESS probability distribution $P(x)$ from simulation. }
\label{Fig_joint_simulation}
\end{figure}

The result, shown in Fig.~\ref{Fig_joint_simulation}, indicates the deviation of $R(x)$ from unity near the temperature discontinuity, signifying the violation of detailed balance.  As mentioned in the previous paragraph, the violation is less significant as $\delta$ becomes smaller (see Fig.~\ref{Fig_joint_simulation}(a)(b)(c) for comparison).  The irreversibility is also more prominent for a larger temperature difference, as demonstrated in Fig.~\ref{Fig_joint_simulation}(d). 
Note that for single-step transitions over identical-temperature regions, Eq.~\eqref{eqn_ratio1} reduces to $F\Delta x/T \cong -\Delta U/T$ (+ higher-order corrections).   According to this prediction, the ratio of joint probabilities $R(x)$ can be represented as
\begin{equation}
	R(x) = \frac{P(x-\delta/2)}{P(x+\delta/2)} \cdot \exp{(-\Delta U/T)} \, .
	\label{eqn_ratio_predicted}
\end{equation}
or equivalently, the ratio of $P(x) \cdot\exp{[U(x)/T]}$ over successive intervals.  With the use of NESS probability distribution extracted from the simulation result, the prediction by Eq.~\eqref{eqn_ratio_predicted} matches well with the joint-probability ratio Eq.~\eqref{eqn_ratio_joint}.  The only exception occurs at the point $x=x_0$, where the forward and backward transitions cross over the temperature discontinuity.  In this case, Eq.~\eqref{eqn_ratio_predicted} cannot be applied.

\subsection{NESS probability distribution}
\label{subsec_SS_P}
\begin{figure}[h]
\includegraphics[width=0.95\textwidth]{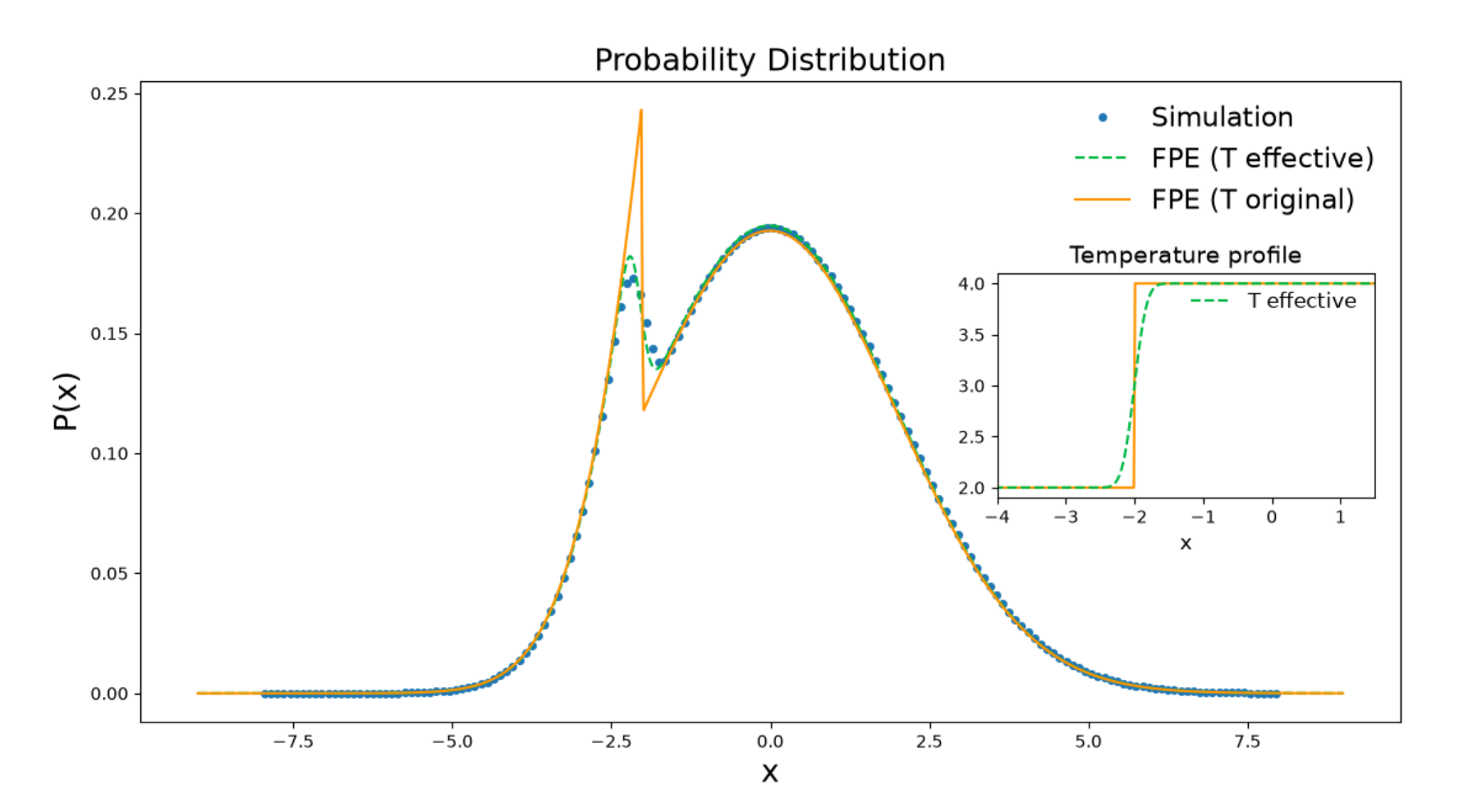}
\includegraphics[width=0.95\textwidth]{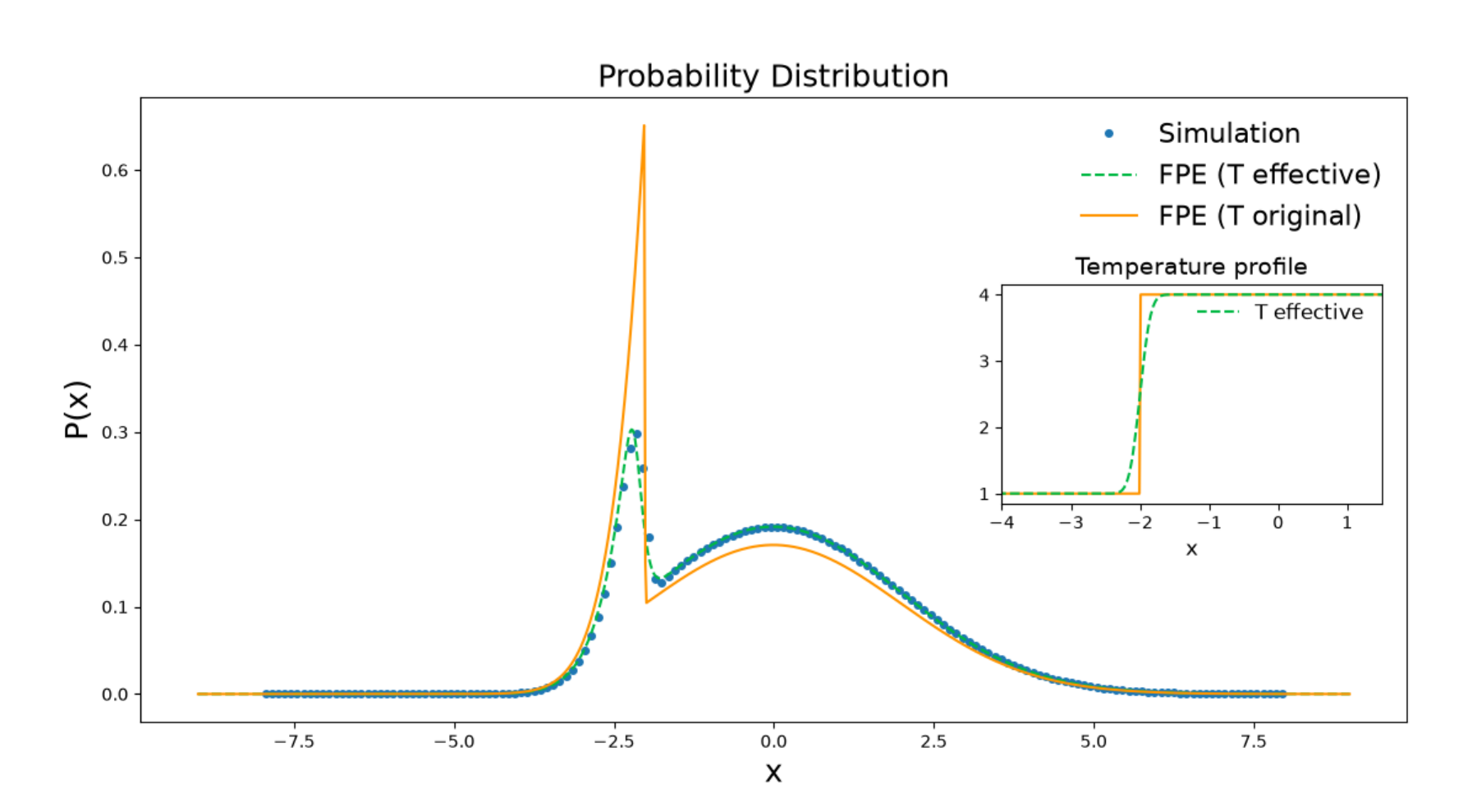}
\centering
\caption{Comparison of NESS probability distributions $P(x)$ for a Brownian particle in a harmonic potential with inhomogeneous temperature profile, for (a) $T_H=4, T_L=2$; (b) $T_H=4, T_L=1$.
	Blue dots represent the simulation results;
	the orange solid line shows the theoretical prediction from Eq.~\eqref{eqn_prob} using the step-function temperature profile;
	the green dashed line represents the modified theoretical prediction using the effective temperature profile from Eq.~\eqref{eqn_eff_temp} (see inset).}
\label{Fig_prob_distribution}
\end{figure}

Note that the FPE analysis Eq.~\eqref{eqn_P_FP} results in the Boltzmann-like behavior in NESS probability distribution over constant-temperature intervals.  
For the open-boundary case, the FPE analysis gives
\begin{equation}
P(x) = \frac{1}{Z} \exp\left(-\int_{-\infty}^{x}\frac{\frac{d}{dx}(U(x)+T(x))}{T(x)}dx \right)\, ,
\label{eqn_prob}
\end{equation}
where Z is the normalization constant.
If the FPE analysis were legitimate, according to Eq.~\eqref{eqn_ratio_predicted}, detailed-balance violation could not be observed in transitions over constant-temperature intervals, as the violation is demonstrated in Fig.~\ref{Fig_joint_simulation}.  A direct comparison between the FPE prediction of the NESS probability distribution and the simulation results, as shown in Fig.~\ref{Fig_prob_distribution}, also reveals some discrepancy.  The discrepancy becomes more significant if the temperature difference gets larger (see Fig.~\ref{Fig_prob_distribution}(b)). 
This deviation is particularly pronounced near the temperature discontinuity at $x=x_0$.  The discrepancy suggests about the inaccuracy of the FPE prediction in the case of discrete-time transitions.

Comparing the results of Figs~\ref{Fig_prob_distribution}(a) and (b), the latter of which with a higher temperature in the hot heat bath, one can observe a reduce of the probability on the high temperature region.  This is because the larger thermal kicks tend to drive the Brownian particle to the cooler side.  Furthermore, one can also note that the simulation result exhibits a larger deviation from the FPE prediction when $T_H$ gets higher.  This implies that the Brownian particle is more prone to move towards the cooler heat bath due to the irreversibility mechanism mentioned in this work.

\subsection{Study in NESS probability current}
\label{subsec_SS_j}
In a one-dimensional open system, the NESS probability current $j(x)$ must be zero everywhere.  This can be understood via the following argument: if one sets up a monitoring station at any arbitrary position, and count the number of events for a random walker to go through this site (the count number gains ``+1'' if the particle is passing from left and ``-1'' vice versa).  And over the long course of a simulation, the total count number can have 3 possible values only: +1, 0, and -1, because the random walker cannot pass the station twice from the same side without returning back in between.  
In our simulation work, we try to numerically examine the NESS probability current via the formula $j(x) = P(x) v(x)$, where $v(x)$ represents the mean local velocity at site $x$\cite{Aurell12}:

\begin{equation}
v(x) \approx \frac{1}{2 \Delta t} \langle [ (x_{t+\Delta t}-x_t) + (x_t-x_{t-\Delta t}) ] |x_t=x\rangle \, .
\label{eqn_v}
\end{equation}

\begin{figure}[h]
\includegraphics[width=0.8\textwidth,height=0.4\textwidth]{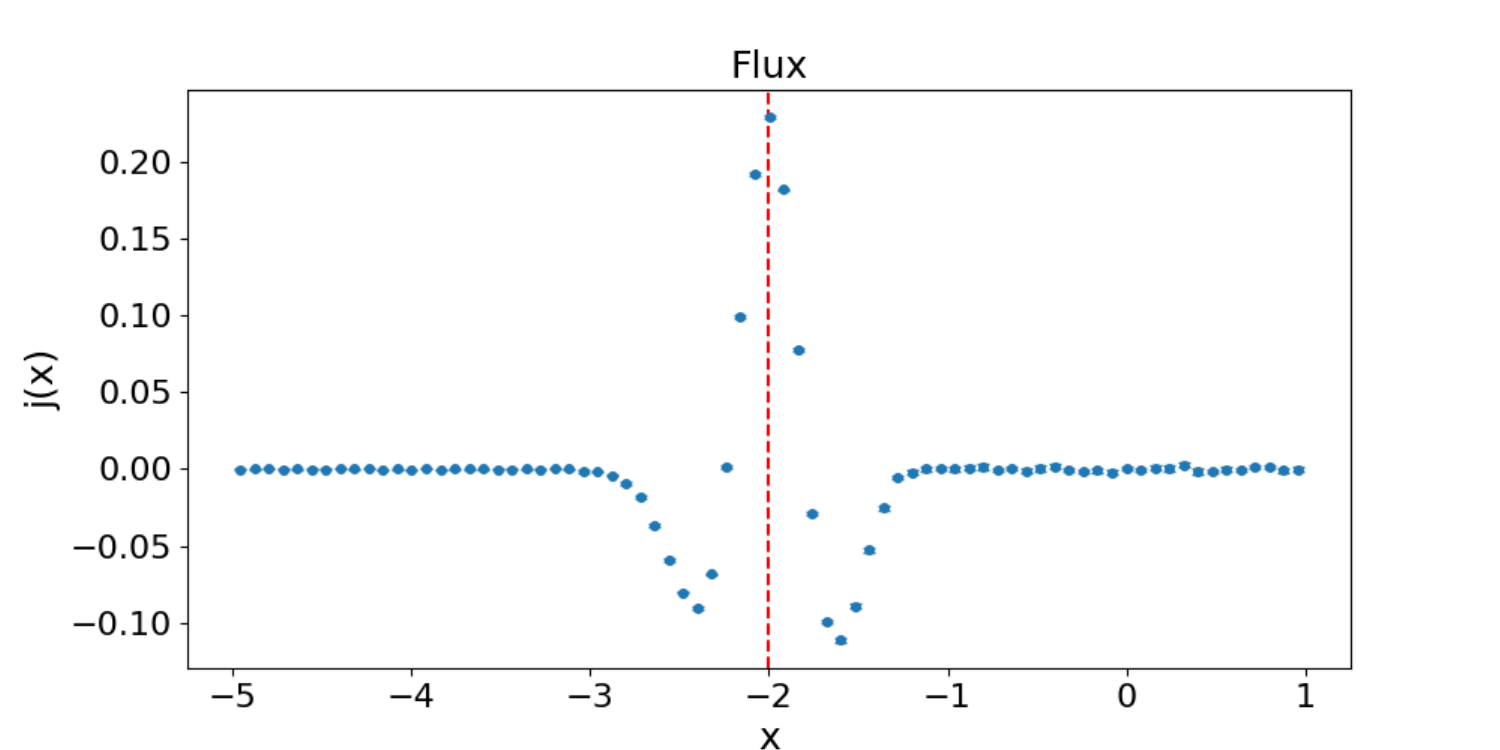}
\centering
\caption{Probability current $j(x)$, evaluated via $j = P(x)v(x)$, where $v(x)$ is obtained through finite difference approximation of particle velocities.  Note that for one-dimensional open-boundary systems, the ideal probability current has to be zero at all positions.}
\label{Fig_flux}
\end{figure}

The result of the computed $j(x)$, as presented in Fig.~\ref{Fig_flux}, exhibits noticeable deviation near the temperature discontinuity.  The deficit is attributed to the fact that aforementioned computing scheme in $v$ and thus $j$ can miss certain events.  To be more specific, in the current model of description, the events that are generated from the diffusion in the high-temperature region possess higher mobility.  And these ``fast'' events, of which the random walker ``bypasses'' the monitoring bin over timestep $\Delta t$, often fail to be counted in the calculation of average velocity at the midpoints of their paths using Eq.~\eqref{eqn_v} (see Fig.~\ref{Fig_illustration}(b)).  The deficit in the calculated probability current is thus more prominent nearby the temperature interface, where the calculated average velocity becomes positive.

\section{Modified Temperature Profile and Probability Distribution}
\label{sec_eff_temp}
The comparison of the simulation result and the FPE prediction in the probability distribution shown in Fig.~\ref{Fig_prob_distribution} shows that the simulation result is less discontinuous in slope across the temperature discontinuity.  Meanwhile, the discrepancy triggers the following physical consideration: in the discrete-time simulation, what is the representative temperature of a Brownian particle crossing a sharp discontinuity at $x_0$.  Nearby the temperature discontinuity, the particle that arrives at a position of consideration at time $t+\Delta t$ may come from positions of either $T_H$ or $T_L$ at time $t$.
The above consideration prompts us to adopt a na\"{i}ve picture with the utilization of an ``effective'' temperature, which differs from the actual step-function profile.  In fact, any arbitrary NESS probability distribution must satisfy, in the mathematical sense, some one-dimensional Fokker-Planck-type equation with the introduction of an effective temperature profile.  By the choice of a proper effective temperature profile, the probability distribution of our discrete-time stochastic dynamics can still be represented by a Fokker-Planck equation.
To fit our simulation result, we propose the following effective temperature profile:
\begin{equation}
T_{\text{eff}}(x) = T_L + (T_H-T_L)\cdot\frac{1}{2}\left[1+\text{erf}\left(\frac{x-x_0}{w}\right)\right]\, ,
\label{eqn_eff_temp}
\end{equation}
with the characteristic width $w$ given by
\begin{equation}
w = \sqrt{2(T_H-T_L)\Delta t}\, .
\label{eqn_eff_temp_width}
\end{equation}
The temperature profile that interpolates between $T_L$ and $T_H$ utilizing the error function seems to serve as an appropriate candidate, as we expect that the smoother transition in temperature can lead to a less sharp behavior in probability density change near the crossing.  The width $w$ is related to the difference in the variances of the diffusion steps between the high- and low-temperature intervals.
Thus $w$ depends on the temperature difference $(T_H-T_L)$ and the timestep $\Delta t$.  Either a larger temperature jump or a longer timestep leads to a broader spread in the effective temperature transition.

In Fig.~\ref{Fig_prob_distribution}, we find that, with the use of the effective temperature profile, the resulting NESS probability distribution of FPE analysis fits quite well with the one obtained by simulation.  In Table \ref{onlytab}, we compare the simulation results with the original FPE prediction as well as the FPE analysis using the effective temperature profile, over various timesteps and temperature settings.  We evaluate the deviation from the simulation probability distribution via the variance, i.e.,
\begin{equation}
{\rm{var}}_i \equiv \sum_j (\delta P_{i,j})^2 \Delta x \, ,
\label{eq_var}
\end{equation}
where the summation is performed over all binned positions, $\delta P_1 = P_{\rm{sim}} - P_{\rm{FP,0}}$ if one compares the deficits between the simulation and original FPE distributions, and $\delta P_2 = P_{\rm{sim}} - P_{\rm{FP,eff}}$ if one compares the simulation and FPE distributions using the effective temperature.  In all our presented trials, the deviations of the effective-temperature FPE prediction from simulations (var$_2$) are at least more than an order less than those derived from the original FPE approach (var$_1$). From Table~\ref{onlytab} one also notices that as $\Delta t$ becomes smaller, the deficit between the simulation result and the original FPE prediction (var$_1$) reduces.  This is due to the fact that the irreversible mechanisms introduced here have a narrower region of influence.  The deficit var$_1$ also reduces if $T_H$ and $T_L$ gets closer, as the nonequilibrium effects become less significant.

\begin{table}
\begin{subtable}[c]{0.5\textwidth}
	\centering
	\begin{tabular}{|l|l|l|}
		\hline
		$\Delta t$ & var$_1$ & var$_2$ \\
		\hline
		0.02 & 1.01E-3 & 4.76E-5\\
		\hline
		0.01 & 7.70E-4 & 3.82E-5\\
		\hline
		0.005 & 5.83E-4 & 3.35E-5\\
		\hline
	\end{tabular}
	\subcaption{}
\end{subtable}
\ \ 
\begin{subtable}[c]{0.5\textwidth}
	\begin{tabular}{|l|l|l|l|}
		\hline
		$T_H$ & $T_L$ & var$_1$ & var$_2$ \\
		\hline
		4 & 1 & 2.87E-2 & 4.82E-4\\
		\hline
		4 & 2 & 7.70E-4 & 3.82E-5\\
		\hline
		6 & 2 & 7.59E-3 & 1.62E-4\\
		\hline
	\end{tabular}
	\subcaption{}
\end{subtable}
\caption{Deviations of theoretical prediction in probability distributions from simulation results.  The deviations (variances) are calculated from Eq.~\eqref{eq_var}, where $(\delta P)_1 = P_{\rm{sim}} - P_{\rm{FP,0}}$ and $(\delta P)_2 = P_{\rm{sim}} - P_{\rm{FP,eff}}$. (a) Deviations for various timesteps $\Delta t$ for the case $T_H=4 ,T_L=2$ ; (b) deviations for various combinations of $T_H$ and $T_L$ for the case $\Delta t = 0.01$.}
\label{onlytab}
\end{table}

The above comparison suggests that one can still derive the distribution of a one-dimensional stochastic simulation over discrete timesteps via an effective FPE analysis.  However, this interpretation may appear to be paradoxical, since we noted earlier that any one-dimensional Fokker-Planck-type equation implies an underlying chain-connecting structure.  And how could the detailed balance be violated at NESS if the underlying structure is chain-like with open boundaries?
We speculate that this paradox may be resolved knowing that time irreversibility is implicitly incorporated through the effective temperature formalism.  To be more specific, an effective temperature at a particular position contains the information of its last-step temperature, and this memory effect is in fact irreversible.

\section{An Estimate for Total Entropy Production}
\label{sec_entropy_prod}
The irreversibility introduced in this work implies an increase in total entropy, and hereby we seek to estimate this quantity.  Through the discrete timestep $\Delta t$, the total entropy production can be written as
\begin{align}
\Delta S_{\rm{total}} &= \sum_{x,x'} P(x', x) \ln \frac{P(x',x)}{P(x,x')} 
= \sum_{x,x'} P(x', x) \ln \frac{P(x) P(x'|x)}{P(x')P(x|x')} \nonumber \\
&= \sum_{x,x'} P(x) P(x'| x) \ln \frac{P(x'|x)}{P(x|x')} \, ,
\label{eqn_total_dS}
\end{align}
where the summation is performed over all pairs of sites $(x, x')$.  According to Eq.~\ref{eqn_forward}, only transitions within the diffusion length scale have substantial contributions in $P(x'|x)$.  Meanwhile, Eq.~\ref{eqn_ratio1} indicates that $\displaystyle \ln\frac{P(x'|x)}{P(x|x')} = O(1)$ for such transitions.  Furthermore, the domain of $x$ where transitions across the temperature interface can occur most likely is in the order of single-step diffusion size $O(\sqrt{\Delta t})$.  As a result, one can deduce that $\Delta S_{\rm{total}} = O(\sqrt{\Delta t})$, and the total entropy production rate is of the order $O(1/\sqrt{\Delta t})$.  


Note that there have been discussions about the average entopy production rate due to temperature inhomogeneity~\cite{Hondou00, Celani12}.  While the work of Ref.~\cite{Hondou00} also yielded a diverging entropy production rate, the diverging behavior is absent in the discussion of Ref.~\cite{Celani12}.  The latter work considered the scenario where temperature is modulated slowly over the spatial dimension, which is distinct from our sharp-interface setup.  While the work in Ref.~\cite{Hondou00} did consider a B\"uttiker-Landauer setup and observe the diverging entropy production, we remark here that their modeling scenario is quite distinct from ours in the following aspects.  First, they considered the full phase-space dynamics, and therefore the stochastic energetics can be fully addressed.  On the other hand, while our overdamped model may have less clear traces on the full version of energetics, we can still observe entropy production through the direct study of time irreversibility.  Second, a fundamental characteristic in our model lies in the fact that the particle gets fully relaxed between random kicks, while in a typical colloidal system, the thermal kicks and relaxation are completely interwined.  Finally, the diverging behavior in entropy production in Ref.~\cite{Hondou00} occurs when the system approaches the overdamped limit, i.e., the mass-to-damping ratio becomes negligible.  As a contrast, the diverging trend in our system is observed when the kick interval approaches the infinitesimal limit. 

\section{Discussion and Summary}
\label{sec_summary}
Throughout this study, we demonstrate the existence of time irreversibility in one-dimensional, open-boundary stochastic dynamics when random thermal kicks occur over discrete time intervals.  Our study originates from a discrete version of the overdamped Langevin equation, where the sources of irreversibility as introduced in Refs.~\cite{Hondou00, Celani12} would be absent.  Yet we still find attributes of irreversibility in this simple model.
Similar to the discussions in other areas of nonequilibrium steady-state stochastic dynamics, the irreversibility observed here can be represented via some ``gyrating'' probability current among positions near the temperature discontinuity (see Fig.~\ref{Fig_illustration}(b)).  The ``hidden'' gyration arises due to the mismatched diffusion distances between the high- and low-temperature adjacent regions.  In terms of probability, the joint probabilities of forward and backward transitions exhibit differences near the temperature discontinuity, while the steady-state probability distribution in position also deviates from the description of the typical FPE analysis.  

Comparing our model with the systems in Ref.~\cite{Hondou00}, one finds a common feature in that the Brownian particle can march across the temperature interface before it completely loses its ``memory'' about the kick.  In Refs.~\cite{Hondou00}, the underdamped nature gives to such memory; in our system, it is the discrete time interval that makes it possible for a Brownian particle to travel across the interface through one single kick.  As the underdamped kinetics in a continuous-time system proceeds towards the overdamped limit, or as the kicking time interval in our model reduces to infinitesimal, the buffer region where the two temperature baths can exchange information via particle diffusion shrinks.  This buffer region can be exhibited through the index of irreversibility $R(x)$ in Fig.~\ref{Fig_joint_simulation}. 

The characteristic width $w$, which has the same order of magnitude as the average diffusion distance over a single timestep, would shrink as the timestep between successive thermal kicks becomes smaller.  Therefore, the irreversible characteristics demonstrated in this work can be observed in experiments if the average diffusion distance due to random kicks over the discrete time interval is large enough to be well resolved.
Note that our model of study does not apply to the conventional colloidal system, whose full irreversible behavior is better described through the usage of underdamped dynamics. 
Our scenario could be realized through a colloidal platform with additional noise feedback control\cite{Jun1, Jun2}.  In such a system setup, there exists a uniform damping background, and on top of that, one can couple the colloidal particle to some external kicks that possess the designed thermal profile.  Other possible candidate systems include the stochastic collisions of granular materials and seismic activities\cite{Puglisi_book, Eshuis10, Hartmann15, Peng16, Serra-Garcia16}.  In particular, the collision time interval for a granular gas can be milliseconds and longer\cite{granular_gas}, which should facilitate the observation.  

It is also worth noting that the shrinking of irreversible characteristic region as $\Delta t \to 0$ does not imply that the accompanied entropy dissipation  due to this algorithm would vanish as well.  According to Eq.~\eqref{eqn_ratio1}, the middle term would actually diverge for finite $\Delta x$ as $\Delta t \to 0$, although this divergence is less significant since the likelihood for finite traveling distance diminishes for infinitesimal kicking timesteps.  
Actually, one can deduce that the average entropy dissipation by the middle term of Eq.~\eqref{eqn_ratio1} is $\displaystyle \frac14 \cdot (T_H + T_L)\cdot (\frac{1}{T_L} - \frac{1}{T_H})= \displaystyle \frac14 \cdot\frac{T_H^2 - T_L^2}{T_L T_H}$ as $\Delta t \to 0$.  For examples, this average entropy dissipation for a crossing transition is 0.0477 for $T_H/T_L=1.1$ and 0.375 for $T_H/T_L=2.0$ (noting that the Boltzmann constant is set to be 1 here).  It can be interpreted as the average information cost per crossing transition for maintaining the sharp temperature interface.  

From the discussion in Sec.~\ref{sec_entropy_prod}, we deduce that the frequency of crossing transitions bears the order $O(1/\sqrt{\Delta t})$.  Therefore, the entropy production rate diverges as the timestep becomes infinitesimal.  In this regard, the square root of timestep is inversely correlated to the ``heat conductivity'' between the two temperature reservoirs due to the particle crossings.  For example, in a colloidal system with the feedback-control protocol, the time interval can be of the order $\Delta t \sim 10 \mu\rm{s}$\cite{Jun18}.  Meanhile, for a 2-dimensional granular media, the mean free path is of the order $\ell \sim 1/(n\sigma)=\sigma\sqrt{3}/2\phi$, where $n$ is the 2-dim number density, $\sigma$ is the particle diameter, and $\phi$ is the packing fraction.  Furthermore, $\Delta t =\ell/v_{\rm rms}$, while $v_{\rm rms} = \sqrt{2 T_g}$ is the particle's root-mean-square speed and $T_g$ is the granular temperature.  For granular particles of $\sigma = $1 mm and $\phi=0.3$, if the granular temperature is $10^4$ mm$^2$/s$^2$, one has $\Delta t \sim $0.015 s.  The evaluation of the kicking time interval can help us to estimate the least amount of information rate to maintain the sharp temperature interface.


Finally, we remark that while we focus on the open-boundary one-dimensional system, the mechanism responsible for the irreversibility remains valid for the closed-loop B\"uttiker-Landauer engine, or higher-dimensional stochastic systems with temperature discontinuity.
In addition to the development in previous discussions originating from underdamped systems, we hope our current study can provide an alternative studying direction in irreversible stochastic thermodynamics.

\begin{acknowledgments}
{\bf Acknowledgment:} This work has been supported by the National Science and Technology Council of Taiwan under grant no.  113-2112-M008-018-MY2 (PYL).
\end{acknowledgments}


\bibliography{jstatphys}

\end{document}